\documentclass[final,10p,times,twocolumn]{elsarticle}

\usepackage{amssymb,amsmath,amsthm,epsfig,bbm,euscript,array,amsfonts}
\usepackage{mathrsfs}
\usepackage{hyperref}
\usepackage[normalem]{ulem}
\usepackage{pstricks}
\usepackage{graphicx}
\begin{document}
\begin{frontmatter}
\title{A gracious exit from the matter-dominated phase in a quantum cosmic phantom model}

\author[csic]{Pedro F. Gonz\'alez-D\'iaz}
\address[csic]{Colina de
los Chopos, Instituto de F\'{\i}sica Fundamental,
Consejo Superior de Investigaciones Cient\'{\i}ficas, Serrano 121,
28006 Madrid, Spain
}
\author[csic,icg]{Alberto Rozas-Fern\'andez}
\ead{a.rozas@iff.csic.es}
\address[icg]{Institute of Cosmology and Gravitation, University of Portsmouth, Dennis Sciama Building, Portsmouth PO1 3FX, United Kingdom
}

\begin{abstract}
The most recent observational constraints coming from Planck, when combined with other cosmological data, provide evidence for a phantom scenario. In this work we consider a quantum cosmic phantom model where both the matter particles and scalar field
are associated with quantum potentials which make the
effective mass associated with the matter particles to vanish at
the time of matter-radiation equality, resulting in a cosmic system where a matter
dominance phase followed by an accelerating expansion is allowed.

\end{abstract}
\begin{keyword}
quantum models \sep dark energy

\end{keyword}

\end{frontmatter}

The problem of dark energy still remains unsolved. Its equation of state (EoS), which is defined as $w=p/\rho$, where $p$ and $\rho$ are the pressure and energy density of dark energy, respectively, could be in the phantom regime ($w<-1$) \cite{Caldwell:1999ew} according to the most recent observational constraints \cite{Ade:2013lta}. Planck latest results \cite{Ade:2013lta} plus WMAP low-\emph{l} polarisation (WP), when combined with Supernova Legacy Survey (SNLS) data, favour the phantom domain at 2$\sigma$ level for a constant $w$
\begin{equation}\label{Planck}
w=- 1.13^{+0.13}_{-0.14} \;(95\%; Planck+WP+SNLS)\;,
\end{equation} while the Union2.1 compilation of 580 Type Ia supernovae (SNe Ia) is more consistent
with a cosmological constant ($w=-1$). If we combine Planck+WP with measurements of $H_{0}$ \cite{Riess:2011yx}, we get for a constant $w$
\begin{equation}\label{Planck2}
w=- 1.24^{+0.18}_{-0.19}\;
\end{equation} which is in tension with $w=-1$ at more than the 2$\sigma$ level. The constant $w$ models are of limited physical interest. If $w\neq-1$ then it is likely to change with time. For a flat universe and for a non-constant $w$ ($w=w_{0}+w_a(1-a)$
\cite{Chevallier:2000qy, Linder:2002et}) the combined data from Planck+WP+$H_{0}$ leads to
\begin{equation}
\label{p1}
w_{0}=- 1.04^{+0.72}_{-0.69}\,
\end{equation} with
a negative $w_{a}$, away from $w=-1$ at just under the 2$\sigma$ level. Furthermore, with the release of
the  first results from Planck \cite{Ade:2013lta}, claims for $w<-1$ at $\geq 2\sigma$ have
been presented, such as \cite{oai:arXiv.org:1310.3828}, which features high-quality
data and a careful analysis including systematic errors
\cite{oai:arXiv.org:1310.3824}. Also, the authors in \cite{Shafer:2013pxa} found that for the SNLS3 and the Pan-STARRS1 survey (PS1 SN)
data sets, the combined SNe Ia + Baryon Acoustic Oscillations (BAO) + Planck data
yield a phantom equation of state at $\sim 1.9\sigma$ confidence. Therefore, we find ourselves in a situation in which we can say \cite{Shafer:2013pxa}, at $2\sigma$ confidence level, that given Planck
data, either the SNLS3 and PS1 data have systematics
that have not been accounted for yet, or the Hubble constant is
below $71$ km/s/Mpc, or else $w<-1$.

The above observational results, in addition to theoretical motivations, are compelling enough to justify the study of the phantom
regime in more depth. Given that the standard cosmological model ($\Lambda$CDM) with $w=-1$ cannot accommodate this scenario, different solutions have been proposed. There are two main approaches. The first one includes a scalar field with a negative kinetic energy term \cite{Caldwell:1999ew} but this leads to violent quantum instabilities \cite{Carroll:2003st, Cline:2003gs}. The second one is more radical and advocates a modification of general relativity. In this modified gravity scenario there are prescriptions that do not have any ghost degree of freedom, such as the the Brans-Dicke type gravity  \cite{Elizalde:2004mq}, the scalar-Einstein-Gauss-Bonnet gravity \cite{Nojiri:2005vv}, and the $F(R)$
gravity \cite{Nojiri:2010wj}. These three proposals are also free of perturbative instabilities but one should also investigate
the corrections to the Newton law, perform the PPN analysis \cite{Will:2005va} etc., in order to ensure that they are consistent with the more accurate solar-system and experimental data.  Furthermore, it was recently realised by some authors
that the most general second order scalar tensor Lagrangian (and thus, ghost-free) that still produces second order equations of motion is the
so-called Horndeski Lagrangian \cite{Horndeski:1974, Deffayet:2011gz, oai:arXiv.org:1105.5723, oai:arXiv.org:hep-th/0609155}, a model that includes four arbitrary functions of the scalar field and its kinetic
energy, and of which Brans-Dicke, Gauss-Bonnet and $F(R)$ are just particular examples.

Alternatively, a theory which is self-consistent and agrees with all the above
observational data \cite{Ade:2013lta} has been proposed \cite{GonzalezDiaz:2006bz, GonzalezDiaz:2006tr, GonzalezDiaz:2008ci}. It is most economical as it only uses general relativity and quantum mechanics without inserting any kind of vacuum
fields or introducing any extra terms in the Hilbert-
Einstein gravitational action. In
such a framework one can get essentially two relevant quantum
solutions both of which can be seen as quantum perturbations to
the de Sitter space \cite{GonzalezDiaz:2006tr}, which is recovered in the classical
limit where $\hbar\rightarrow 0$. It has also been shown that out
of these two possible solutions only one of them satisfies the
second law of thermodynamics \cite{GonzalezDiaz:2008ci}, and hence is physically
meaningful. It corresponds to a phantom universe \cite{Caldwell:1999ew} but does not show any quantum instability \cite{Carroll:2003st, Cline:2003gs} nor the sort of inconsistency coming from having a negative kinetic
term for the scalar field - in fact, these models do not actually
contain any scalar or other kinds of vacuum fields in their final
equations and do not show neither a future
singularity (Big Rip) \cite{Caldwell:1999ew, Caldwell:2003vq} nor classical violations of the energy conditions. It is for these reasons that such a cosmic model has also been denoted as \cite{GonzalezDiaz:2006tr, GonzalezDiaz:2008ci} {\it benigner} phantom model.

On the other hand, in Ref.\ \cite{Amendola:2006qi} (see also \cite{Gomes:2013ema}) it was shown that it is impossible to find a sequence
of matter and scaling acceleration for any scaling Lagrangian
which can be approximated as a polynomial because a scaling
Lagrangian is always singular in the phase space so that either
the matter-dominated era is prevented or the region with a viable
matter is isolated from that where the scaling acceleration
occurs. Such as it happens with
other aspects of the current accelerating cosmology, the problem
is to some extend reminiscent of the difficulty initially
confronted by earliest inflationary accelerating models \cite{Guth:1980zm} which
could not smoothly connect with the following
Friedmann-Robertson-Walker (FRW) decelerating evolution \cite{Padmanabhan:2002ji}. As is
well known, such a difficulty was solved by invoking the new
inflationary scenario \cite{Linde:1981mu}. In fact, the problem posed in \cite{Amendola:2006qi} for
dark energy can be formulated by saying that a previous
decelerating matter-dominated era cannot be followed by an
accelerating universe dominated by dark energy and it is in this
sense that it can be somehow regarded as the time-reversed version
of the early inflationary exit difficulty.  Ways out from this problem required assuming either a
sudden emergence of dark energy domination or a cyclic occurrence
of dark energy, both assumptions being quite hard to explain and
implement. The aim of this work is to show that in the {\it benigner} phantom model \cite{GonzalezDiaz:2006tr, GonzalezDiaz:2008ci}
 such problems are no longer present due to the
quantum characteristics that can be assigned to particles and
radiation in this model.

If we apply the real part of the Klein-Gordon wave equation to a
quasi-classical wave function $R\exp(iS/\hbar)$, where the
probability amplitude $R$ ($P=|R|^2$) and the action $S$ are real
functions of the relativistic coordinates, and define the classical energy
$E=\partial S/\partial t$ and momentum $p=\nabla S$,
we can write the modified Hamilton-Jacobi equation
\begin{equation}\label{modHJ}
E^2-p^2+\tilde{V}_{Q}^2=m_0^2,
\end{equation}
where $m_0$ is the rest mass of the involved particle and $\tilde{V}_{Q}$
is a relativistic quantum potential,
\begin{equation}\label{quantumpot}
\tilde{V}_{Q}^2= \frac{\hbar^2}{R}\left(\nabla^2 R-\frac{\partial^2
R}{\partial t^2}\right) ,
\end{equation}
which should be interpreted according to Bohm's idea \cite{Bohm:1951xw} as
the hidden quantum potential that accounts for precisely
defined unobservable relativistic variables whose effects would
physically manifest in terms of the indeterministic behaviour
shown by the given particles. From Eq.\ (\ref{modHJ}) it immediately
follows that $p=\sqrt{E^2+\tilde{V}_{Q}^2-m_0^2}$. Thus, since
classically $p=\partial \tilde{L}/\partial[\dot{q(t)}]$ (with $\tilde{L}$ being
the Lagrangian of the system and $q$ the spatial coordinates, which
depends only on time $t$, $q\equiv q(t)$), we have for the
Lagrangian
\begin{equation}
\tilde{L}=\int d\dot{q}p=\int dv\sqrt{\frac{m_0^2}{1-v^2}+M^2} ,
\end{equation}
in which $v=\dot{q}$ and $M^2=\tilde{V}_{Q}^2-m_0^2$. In the classical limit $\hbar\rightarrow0$, $\tilde{V}_{Q}\rightarrow0$, and hence we are just left with the classical relativistic Lagrangian for a particle with rest mass $m_0$.

We start with an action integral that contains all the ingredients
of our model. Such an action is a generalisation of the one used
in \cite{Amendola:2006qi} which contains a time-dependent coupling between dark
energy and matter and leads to a general Lagrangian that admits
scaling solutions formally the same as those derived in \cite{Amendola:2006qi}.
Setting the Planck mass to unity, our Lorentzian action reads
\begin{eqnarray}\label{action1}
&&S=\int d^4 x\sqrt{-g}\left[R+p(X,\phi)\right]\nonumber\\
&&+S_m\left[\psi_i, \xi, m_i(\tilde{V}_{Q}), \phi, g_{\mu\nu}\right]
+ST\left(K,\psi_i, \xi\right) ,
\end{eqnarray}
where $g$ is the determinant of the four-metric, $p$ is a
generically non-canonical general Lagrangian for the dark energy
scalar field $\phi$ with kinetic term
$X=g^{\mu\nu}\partial_{\mu}\phi\partial_{\nu}\phi$, formally the
same as the one used in \cite{Amendola:2006qi}, $S_m$ corresponds to the
Lagrangian for the matter fields $\psi_i$, each with mass $m_i$,
which is going to depend on the quantum potential $\tilde{V}_{Q}$ in a
way that will be made clear in what follows, so as on the
time-dependent coupling $\xi$ of the matter field to the dark
energy field $\phi$. The term $ST$ denotes the surface term which
generally depends on the trace on the second fundamental form $K$,
the matter fields $\psi_i$ and the time-dependent coupling
$\xi(t)$ between $\psi_i$ and $\phi$ for the following reasons.

We first of all point out that in the theory being considered the
coupling between the matter and the scalar fields can generally be
regarded to be equivalent to a coupling between the matter fields
and gravity plus a set of potential energy terms for the matter
fields. In fact, if we restrict ourselves to this kind of
theories, a scalar field $\phi$ can always be mathematically
expressed in terms of the scalar curvature $R$ \cite{Nojiri:2005am}. More
precisely, for the scaling accelerating phase we shall consider a
quantum dark energy model (see \cite{Bohm:1951xw} and \cite{GonzalezDiaz:2006tr, GonzalezDiaz:2008ci}) in which the
Lagrangian for the field $\phi$ vanishes in the classical limit
where the quantum potential is made zero; i.e. we take
$p=L=-V(\phi)\left(E(x,k)-\sqrt{1-\dot{\phi}^2}\right)$, where
$V(\phi)$ is the density of potential energy associated to the field $\phi$ and $E(x,k)$ is the elliptic
integral of the second kind, with $x=\arcsin\sqrt{1-\dot{\phi}^2}$
and $k=\sqrt{1-V_{Q}^2/V(\phi)^2}$, and the overhead dot $\dot{}$
means derivative with respect to time. We do not expect
$\tilde{V}_{Q}$ to remain constant along the universal expansion
but to increase like the volume of the universe ${\rm V} \propto a^{3}$ does. It
is the quantum potential density $V_{Q}=\tilde{V}_{Q}/{\rm V}$
appearing in the Lagrangian $L$ what should be expected to remain constant at
all cosmic times. Using then a potential
energy density for $\phi$ and the quantum medium [note that
the quantum potential energy density becomes constant \cite{GonzalezDiaz:2006tr, GonzalezDiaz:2008ci} (see
later on)], we have for the energy density and pressure,
$\rho\propto X(HV_{Q}/\dot{H})^2 =p(X)/w(t)$, with $H\propto \phi
V_{Q}+H_0$, $\dot{H}\propto \sqrt{2X}V_{Q}$, where $H_0$ is
constant. For the resulting field theory to be finite, the
condition that $2X=1$ (i.e. $\phi=C_1+t$) had to be satisfied \cite{GonzalezDiaz:2006tr, GonzalezDiaz:2008ci},
and from the Friedmann equation the scale factor ought to be
given by $a(t)\propto\exp\left(C_2 t+C_3 t^2\right)$, with $C_1$,
$C_2$ and $C_3$ being constants. It follows then that for at least
a flat space-time, we generally have $R\propto 1+\alpha\phi^2$
(where $\alpha$ is another constant and we have re-scaled time) in
that type of theories, and hence the matter fields - scalar field
couplings, which can be generally taken to be proportional to
$\phi^2\psi_i^2$, turn out to yield $\xi R\psi_i^2 - K_0\psi_i^2$,
with $K_0$ again a given constant. The first term of this
expression corresponds to a coupling between matter fields and
gravity which requires an extra surface term, and the second one
ought to be interpreted as a potential energy term for the matter
fields $V_i\equiv V(\psi_i)\propto\psi_i^2$. In this way, for a
general theory that satisfied the latter requirement, the action
integral (\ref{action1}) should be rewritten as
\begin{eqnarray}\label{action2}
&&S=\int d^4 x\sqrt{-g}\left[R(1-\xi\psi_i^2)+p(X,\phi)\right]\nonumber\\
&&+S_m\left[\psi_i, V_i, m_i(V_{Q}), g_{\mu\nu}\right]\nonumber\\
&& - 2\int d^3 x\sqrt{-h}{\rm Tr}K(1-\xi\psi_i^2) ,
\end{eqnarray}
in which $h$ is the determinant of the three-metric induced on the
boundary surface and it can be noticed that the scalar field
$\phi$ is no longer involved in the matter Lagrangian. We
specialise now in the minisuperspace that corresponds to a flat
FRW metric in conformal time $\eta=\int
dt/a(t)$
\begin{equation}
ds^2=-a(\eta)\left(-d\eta^2+a(\eta)^2 d{\rm x}^2\right) ,
\end{equation}
with $a(\eta)$ the scale factor. There are two choices for $\xi$ of particular interest. The first one is $\xi=0$, i.e., there is no coupling of the field with the spacetime scalar curvature. This is called \emph{minimal coupling}. With this choice, we do not have the most general equation of motion for a scalar field in a curved spacetime background. The second choice is the one we shall take, $\xi=1/6$, known as the conformal coupling. This is a case of great interest in cosmological scenarios given that the FRW metrics are conformally flat. Therefore, if we assume a
time-dependence of the coupling such that it reached the value
$\xi(\eta_c)=1/6$ at the time of matter-radiation equality $\eta_c$ and choose
suitable values for the arbitrary constants entering the above
definition of $R$ in terms of $\phi^2$, then the action at this
time of equality would reduce to
\begin{eqnarray}\label{S}
&&S=\frac{1}{2}\int
d\eta\left[a'^2-\sum_i(\chi_i'^2-\chi_i^2)\right.\nonumber\\
&&+\left.a^4\left(p(X,\phi)+\sum_i m_i(V_{Q})^2\right)\right] ,
\end{eqnarray} where the prime $'$ denotes derivative with respect to conformal
time $\eta$ and $X=\frac{1}{2a^2}(\phi')^2$. Clearly, the fields
$\chi_i$ would then behave like though if they formed a collection
of conformal radiation fields were it not by the presence of the
nonzero mass terms $m_i^2$ also at the time of matter-radiation equality. If for
some physical cause the latter mass terms could all be made to
vanish at this time of equality, then all matter fields would
behave like though they were a collection of radiation fields
filling the universe at around this equality time and there
would not be the disruption of the evolution from a
matter-dominated era to a stable accelerated scaling solution of
the kind pointed out in \cite{Amendola:2006qi}, but the system smoothly would enter the
accelerated regime after a given brief interlude where the matter
fields behave like pure radiation. In what follows we shall show
that in the quantum scenario considered above such a
possibility can actually be implemented.

At the end of the day, any physical system always shows the actual
quantum nature of its own. One of the most surprising implications
tough by dark energy and phantom energy scenarios is that the
universal system is not exception on that at any time or value of
the scale factor. Thus, we shall look at the particles making up
the matter fields in the universe as satisfying the Klein-Gordon
wave equation \cite{comment} for a Bohmian quasi-classical wave function \cite{Bohm:1951xw}
$\Psi_i=R_i\exp(iS_i/\hbar)$, where we have restored an explicit
Planck constant, $R_i$ is the probability amplitude for the given
particle to occupy a certain position within the whole homogeneous
and isotropic space-time of the universe, as expressed in terms of
relativistic coordinates, and $S_i$ is the corresponding classical
action also defined in terms of relativistic coordinates.

The quantum potential for each particle is given by (see Eq.\ (\ref{quantumpot}))
\begin{equation}
\tilde{V}_{Qi}=\hbar\sqrt{\frac{\nabla^2 R_i-\ddot{R}_i}{R_i}} ,
\end{equation}
that should also satisfy the continuity equation (i.e. the
probability conservation law) for the probability flux, $J=\hbar\;
{\rm Im}(\Psi^*\nabla\Psi)/(m{\rm V})$ (with ${\rm V}\propto a^3$
the volume), stemming from the imaginary part of the expression
that results by applying the Klein-Gordon equation to the wave
equation $\Psi$. Thus, if the particles are assumed to move
locally according to some causal laws \cite{Bohm:1951xw}, then the classical
expressions for $E_i$ and $p_i$ will be locally satisfied.
Therefore we can now interpret the cosmology resulting from the
above formulae as a classical description with an extra
quantum potential, and average the modified Hamilton-Jacobi equation
\begin{equation}\label{modHJ2}
E_{i}^2-p_{i}^2+\tilde{V}_{Qi}^2=m_{0i}^2,
\end{equation} with a probability
weighting function for which we take $P_i=|R_i|^2$, so that
\begin{eqnarray}
&&\int\int\int dx^3
P_i\left(E_i^2-p_i^2+\tilde{V}_{Qi}^2\right)\nonumber\\ &&=\langle
E_i^2\rangle_{\rm av}-\langle p_i^2\rangle_{\rm av}+\langle
\tilde{V}_{Qi}^2\rangle_{\rm av} = \langle m_{0i}^2\rangle_{\rm av},
\end{eqnarray}
with the averaged quantities coinciding with the corresponding
classical quantities and the averaged total quantum potential
squared being given by $\langle \tilde{V}_{Qi}^2\rangle_{\rm
av}=\hbar^2\left(\langle\nabla^2 P\rangle_{\rm
av}-\langle\ddot{P}\rangle_{\rm av}\right)$.

It is worth noticing that in the above scenario the velocity of
the matter particles should be defined to be given by
\begin{equation}
\langle v_i\rangle_{\rm av}=\frac{\langle p_i^2\rangle_{\rm
av}^{1/2}}{\left(\langle p_i^2\rangle_{\rm av}+\langle
m_{0i}^2\rangle_{\rm av}-\langle \tilde{V}_{Qi}^2\rangle_{\rm
av}\right)^{1/2}} .
\end{equation}
It follows that in the presence of a quantum potential, a
particle with nonzero rest mass $m_{0i}\neq 0$ can behave like
though if was a particle moving at the speed of light (i.e. a
radiation massless particle) provided $\langle
m_{0i}^2\rangle_{\rm av}=\langle \tilde{V}_{Qi}^2\rangle_{\rm av}$. Thus,
if we introduce an effective particle rest mass $m_{0i}^{{\rm
eff}}=\sqrt{\langle m_{0i}^2\rangle_{\rm av}-\langle
\tilde{V}_{Qi}^2\rangle_{\rm av}}$, then we get that the speed of light
again corresponds to a zero effective rest mass. It has been
noticed \cite{GonzalezDiaz:2006tr, GonzalezDiaz:2008ci}, moreover, that in the cosmological context the
averaged quantum potential defined for all existing radiation
in the universe can be expressed
in terms of a scalar field $\phi$, and would actually make up our
scaling dark energy solution. At the time of matter-radiation equality, that idea
should actually extend in the present formalism to also encompass
in an incoherent way, together with the averaged quantum
potential for CMB radiation, the averaged quantum potential
for matter particles, as a source of dark energy. On the other
hand, as it has been pointed out above as well as in \cite{GonzalezDiaz:2006tr, GonzalezDiaz:2008ci}, the quantum
potential ought to depend on the scale factor $a(t)$ in such a way
that it steadily increases with time, being the quantum energy
density satisfying the above continuity equation what keeps
constant along the whole cosmic evolution.

Assuming the mass $m_i$ appearing in the action (\ref{S}) is an
effective particle mass, it turns out that the onset of dark
energy dominance would then be precisely at the time of matter-radiation equality
when $\langle \tilde{V}_{Qi}^2\rangle_{\rm av}\equiv \langle
\tilde{V}_{Qi}(a)^2\rangle_{\rm av}$ reached a value which equals
$\langle m_{0i}^2\rangle_{\rm av}$ and all the matter fields
behaved in this way like a collection of radiation fields which
are actually irrelevant to the issue of the incompatibility of the
previous eras with a posterior stable accelerated current regime.
In this case, the era of matter dominance can be smoothly followed
by the current accelerated expansion where all matter fields would
effectively behave like though if they cosmologically were
tachyons. This interpretation would ultimately amount to the
unification of dark matter and dark energy, as the dark energy
model being dealt here with is nothing but a somehow quantised
version of tachyon dark energy \cite{Gibbons:2002md}, so that one should expect both
effective tachyon matter and tachyon dark energy to finally decay
to dark matter, so providing a consistent solution to the cosmic
coincidence problem.

Now, from our action integral (\ref{S}) one can derive the equation of
motion for the field $\phi$; that is (see also \cite{Piazza:2004df} and
\cite{Tsujikawa:2004dp})
\begin{equation}
\ddot{\phi}\left(p_X+2Xp_{XX}\right) +3Hp_X\dot{\phi}+
2Xp_{X\rho}-p_{\phi}= \frac{\delta S}{a^3 \delta\phi} ,
\end{equation}
where we have restored the cosmic time $t$, using the notation of
Refs. \cite{Amendola:2006qi}, \cite{Piazza:2004df} and \cite{Tsujikawa:2004dp}, so that a suffix $X$ or $\phi$ denotes a
partial derivative with respect to $X$ or $\phi$, respectively,
and now the last coupling term is time-dependent. Note that if we
confine ourselves to the theory where $a(t)$ accelerates in an
exponential fashion and $\dot{\phi}^2=1$ then the first term of
this equation would vanish. Anyway, in terms of the energy density
$\rho$ for the scalar field $\phi$ the above general equation
becomes formally the same as that which was derived in \cite{Amendola:2006qi}
\begin{equation}
\frac{d\rho}{dN}+3(1+w)\rho=-Q\rho_m \frac{d\phi}{dN} ,
\end{equation}
with $\rho_m$ the energy density for the matter field, $N=\ln a$,
and $Q=-\frac{1}{a^3\rho_m}\frac{\delta S_m}{\delta\phi}$. We can
then derive the condition for the existence of scaling solutions
for time-dependent coupling which, as generally the latter two
equations are formally identical to those derived in \cite{Amendola:2006qi}, is the
same as that was obtained by these authors. Hence, we have the
generalised master equation for $p$ \cite{Amendola:2006qi}
\begin{equation}
\left[1+\frac{2dQ(\phi)}{\lambda Q^2
d\phi}\right]\frac{\partial\ln p}{\partial\ln X}
-\frac{\partial\ln p}{\lambda Q\partial\phi}=1 ,
\end{equation}
whose solution was already obtained in \cite{Amendola:2006qi} to be
\begin{equation}\label{Lag}
p(X,\phi)=XQ(\phi)^2 g\left(XQ(\phi)^2
e^{\lambda\kappa(\phi)}\right)
\end{equation}
where $g$ is an arbitrary function, $\lambda$ is a given function
of the parameters of the equations of state for matter and $\phi$
and the energy density for $\phi$, being $\kappa=\int^{\phi}
Q(\xi)d\xi$ (see \cite{Amendola:2006qi}). In the phase space we then have an
equation-of-state effective parameter for the system $w_{\rm
eff}=-1-\frac{2\dot{H}}{3H^2}=gx^2+z^2/3$, with $H$ the Hubble
parameter and $x$ and $z$ respectively being
$x=\dot{\phi}/(\sqrt{6}H)$ and $z=\sqrt{\rho_{\rm rad}/(3H^2)}$.
At the time of equality where we have just radiation ($z\neq 0$
and $\rho_m=\rho_{\rm rad}$) the effective equation of state is
\cite{Amendola:2006qi} $w_{\rm eff}=1/3$. Hence at the time of equality interval we
can only have radiation, neither matter or accelerated expansion
domination, just the unique condition that would allow the
subsequent onset of the accelerated expansion era where conformal
invariance of the field $\chi$ no longer holds.

Thus, in the considered quantum cosmic phantom model, a previous
matter-dominated phase can be evolved first into a radiation phase
at a physical regular short stage which is then
destroyed to be finally followed by the required new, independent
phase of current accelerating expansion. This conclusion can be
more directly drawn if one notices that there is no way by which
the general form of the Lagrangian (\ref{Lag}) can accommodate the
Lagrangian final form $L\equiv p=f(a,\dot{a})\dot{\phi}^2
V_{Q}^2$ which characterises quantum dark energy models whose
pressure $p$ vanishes in the limit $V_{Q}\rightarrow 0$. Hence,
at least these models can be taken to be counter
examples to the general conclusion that current dark energy and
modified gravity models (see however \cite{Capozziello:2006dj}) are incompatible
with the existence of a previous matter-dominated phase, as
suggested in \cite{Amendola:2006qi}.

We finally notice, moreover, that the kind of quantum dark
energy theory providing the above counter example is one which
shows no classical analog (i.e. the Lagrangian, energy density and
pressure are all zero in the classical limit $\hbar\rightarrow 0$)
and is thereby most economical of all. Thus, the above conclusion
can also be stated by saying that, classically, a previous phase
of matter dominance is always compatible with the ulterior
emergence of a dominating phase made up of "nothing". In this way,
similarly to as the abrupt, nonphysical exit of the old
inflationary problem was circumvented by introducing \cite{Linde:1981mu} a scalar
field potential with a flat plateau leading to a "slow-rollover"
phase transition, the abrupt disruption of the scaling phase after
matter dominance can be also avoided by simply considering a
vanishing scalar field potential that smooths the transition and
ultimately makes it to work.

\section*{Acknowledgments}{ARF is supported by the `Fundaci\'on Ram\'on
Areces' and Ministerio de Econom\'ia y Competitividad (Spain) through project number FIS2012-38816.}


\begin{thebibliography}{00}
\bibitem{comment} Strictly speaking, observable baryonic matter fields should be described
by the Dirac equation rather than the Klein-Gordon equation.
However, for our present purposes the use of the latter equation
will suffice. On the other hand, even before coincidence time,
most of the matter content had to be in the form of dark matter, a
material whose nature is still unknown.
\end{thebibliography}
\end{document}